\documentclass{article}

\usepackage{arxiv}

\usepackage[utf8]{inputenc} 
\usepackage[T1]{fontenc}    
\usepackage{hyperref}       
\usepackage{url}            
\usepackage{booktabs}       
\usepackage{amsfonts, cleveref}       
\usepackage{nicefrac}       
\usepackage{microtype}      
\usepackage{lipsum}		
\usepackage{graphicx,multirow, subcaption}
\usepackage[numbers,sort&compress]{natbib}
\usepackage{doi,comment}

\title{Achieving passive daytime radiative cooling via TiO$_2$/PDMS coating}

\author{ \href{http://orcid.org/0000-0002-6894-8186}{\includegraphics[scale=0.06]{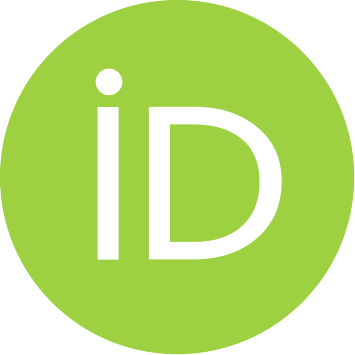}\hspace{1mm}Karthik Sasihithlu\thanks{Corresponding author: ksasihithlu@iitb.ac.in}}, \href{http://orcid.org/0000-0001-6002-8193}{\includegraphics[scale=0.06]{orcid.pdf}\hspace{1mm}Sreerag Sundaram}, \href{http://orcid.org/0000-0002-1900-7603}{\includegraphics[scale=0.06]{orcid.pdf}\hspace{1mm}Bhrigu Rishi Mishra}, Nithin Jo Varghese \\ 
	Department of Energy Science and Engineering\\
	Indian Institute of Technology Bombay\\
	Mumbai -- 400076. India. \\
}

\renewcommand{\headeright}{Technical Report}
\renewcommand{\undertitle}{Technical Report}
\renewcommand{\shorttitle}{TiO$_2$/PDMS coating for PDRC applications}

\hypersetup{
pdftitle={A template for the arxiv style},
pdfsubject={q-bio.NC, q-bio.QM},
pdfauthor={David S.~Hippocampus, Elias D.~Striatum},
pdfkeywords={First keyword, Second keyword, More},
}

\begin{document}
\maketitle

\begin{abstract}

The exponential growth in population and the increasing global temperature trickles down to an explosive demand in cooling and refrigeration. The vicious cycle of carbon footprint generation by these cooling devices can be broken by a mechanism of passive cooling. This study outlines research undertaken with the aim to design such materials -- exhibiting high reflectance in the solar spectrum and high emission in the atmospheric transparency window of 8--13~$\mu$m. The Monte Carlo (MC) method is used to simulate light propagation in a composite material aiding the design of metamaterials with these specific thermo-optical properties. A TiO$_2$/PDMS coating is fabricated to obtain >~91~\% solar reflectivity and >~75~\% emissivity in the atmospheric transparency window. This translates to cooling the coated body by 4-9~$^\circ$C below the ambient under peak solar irradiation in Mumbai, India. The facile fabrication process supplemented with the potential versatility of this coating shows promise to attain passive daytime radiative cooling on a commercial scale.

\end{abstract}


\section{Introduction}
\label{sec:intro}

Estimates suggest that emissions from the "cooling" sector alone would exhaust the carbon budget that targets limiting global warming to 1.5~$^\circ$C above pre-industrial levels. Thus, this calls for a joint and continuous action to curb and bring about significant improvements in this technology. Currently, approximately 3.6 billion units of cooling appliances are in use and this is expected to reach 14 billion units by 2050. A move to the best cooling technologies alone can reduce global cumulative emissions by 38--60 gigatonnes of CO$_2$e by 2030. Making cooling less polluting, therefore, becomes vital in the race to mitigate global warming \cite{CERep, UNEP_2}.

A coating, inspired from the Saharan silver ant \cite{Shi}, mimics its thermoregulatory process to achieve passive daytime radiative cooling (PDRC). The desired coating should be: (a) highly reflective in the solar spectrum (300--2500~nm) and, (b) highly emissive in the atmospheric transparency window (8--13~$\mu$m). Theoretically, it has the potential to reduce the temperature of a surface, passively, by 60~$^\circ$C \cite{chen}. Several materials designed to attain PDRC have demonstrated sub-ambient cooling by 5–7~$^\circ$C viz. Mandal et al.'s air-filled polymer \cite{mondal}, Huang and Ruan's double-layer coating \cite{huang}, and Raman et al.'s seven layered metal/metal oxide composite \cite{raman}. More recently, a study of a ZrO$_2$--PDMS mixture reported having attained a surface temperature reduction of about 10.9~$^\circ$C under a solar irradiation of 895 W m$^{-2}$ \cite{zhang}.

In this study, we build a model based on the Monte Carlo method that allows us to design metamaterials with desired properties. A coating of titania (TiO$_2$) infused in a polymer matrix is observed to exhibit properties that help attain PDRC. This coating is fabricated and experimentally tested to ascertain the claims made by the computational model. 

The PDMS/TiO2 composite is highly stable, hydrophobic, and biocompatible \cite{Bhrigu}. TiO$_2$ is known to absorb in the ultraviolet regime -- mainly in the UVC (100–280 nm) and UVB (280–315 nm) bands. However, incident radiation in the UVC and UVB bands is almost completely blocked by the Earth’s ozone layer and atmosphere, lowering chances of photo-degradation upon exposure to the environment \cite{Bhrigu}. Studies also indicate the ability of PDMS to suppress photo-degradation of TiO$_2$ particles even further. This has enabled the use of the PDMS--TiO$_2$ composite for various outdoor applications making it a viable candidate for PDRC coatings.

\section{Computational design}
\label{sec:headings}

\begin{figure}[htb]
    \centering
    
    \subfloat[RI of PDMS \cite{PDMS_RI1,PDMS_RI2}]{\includegraphics[width=0.495\textwidth]{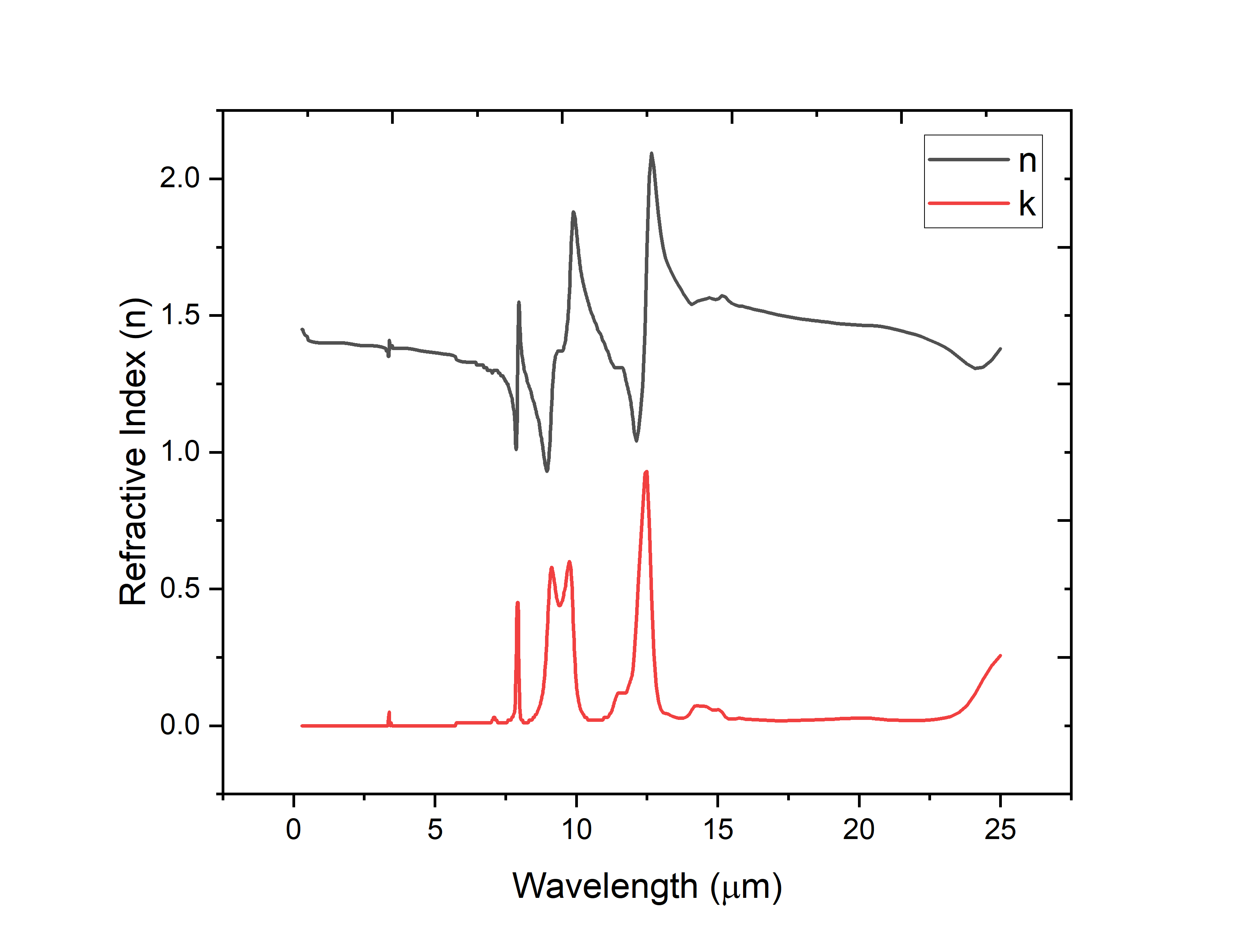} \label{fig:PDMS_RI}} 
    \subfloat[RI of TiO$_2$ \cite{TiO2_RI}]{\includegraphics[width = 0.495\textwidth]{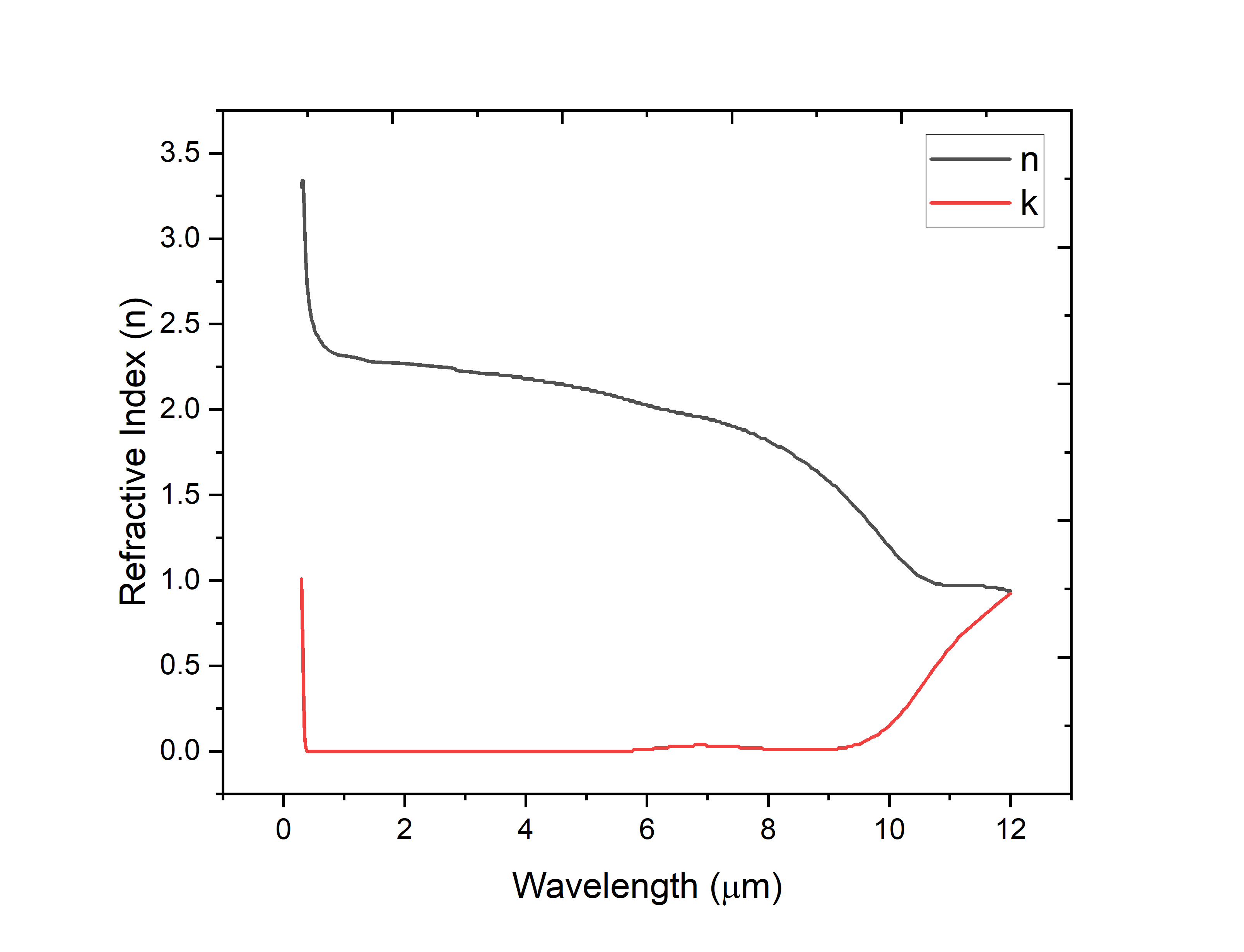} \label{fig:TiO2_RI}}
    
    \subfloat[Author's work \cite{peoples}]{\includegraphics[width=0.43\textwidth]{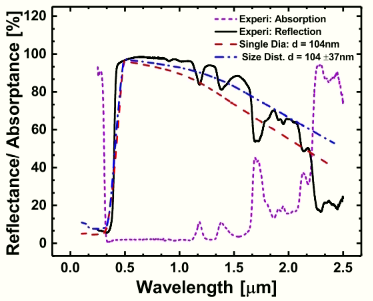}\label{fig:peoples_main}} 
    \subfloat[Comparison with MC]{\includegraphics[width = 0.51\textwidth]{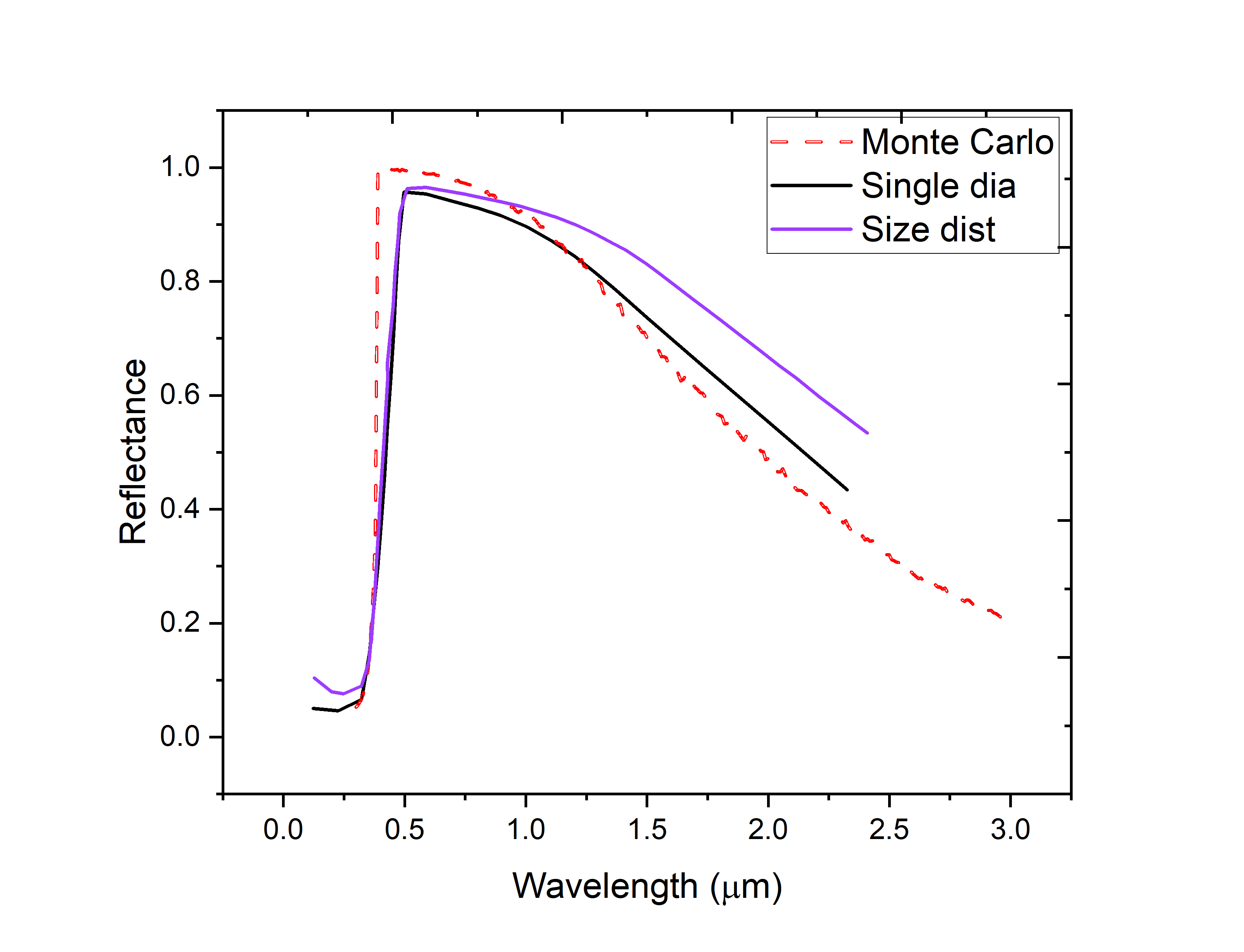} \label{fig:verif_peoples}} 
    
    \caption{(a) and (b) Input parameters used in the model. (c) and (d) Verification of the Monte Carlo method.}
    \label{fig:RI&verification}
    
\end{figure}

The metamaterial design consists of a matrix with spherical inclusions, analogous to extracellular fluid and the corpuscular components in a human tissue. Numerical simulations of the designed composite material are based on light propagation in biological tissue. The Monte Carlo method for multi-layered tissues, designed by Wang and Jacques \cite{Wang_MCML}, is used to solve the Radiative Transfer Equation. This stochastic method models steady–state photon propagation in a material whose properties are derived from Mie theory. The refractive index (RI) data used for calculating the material parameters is depicted in \cref{fig:PDMS_RI,fig:TiO2_RI} \cite{PDMS_RI1,PDMS_RI2,TiO2_RI}. \Cref{fig:verif_peoples} compares the MC result, for a uniform particle distribution, with that modelled by Peoples et al. (\cref{fig:peoples_main}) \cite{peoples} – verifying the effectiveness of the MC method.\\ This framework is used to design a composite comprising titania (TiO$_2$) microspheres infused in a PDMS (polydimethylsiloxane) matrix. Simulation results, published in our previous work \cite{Bhrigu}, shows significant improvement in performance over others reported in literature.

\section{Experimental validation}
\label{sec:expt_valid}

\subsection{Materials, synthesis and characterisation}
PDMS (Part A, SYLGARD\textsuperscript{TM} 184 Silicone Elastomer Kit, Dow) and its curing agent (Part B) are mixed in a 10:1 ratio (by weight) and stirred. n-Hexane (EMPLURA\textsuperscript{\textregistered}, Sigma-Aldrich) is added to this mix in a 2:1 weight ratio \cite{hexane} (PDMS+curing agent : hexane) and stirred while sonicating, until a clear solution is obtained. TiO$_2$ powder (titanium (IV) oxide, anatase powder, -325 mesh, Sigma-Aldrich), weighed to attain a volume fraction of 10~\%, is then added and the resultant solution is sonicated for five minutes. 

This was covered with a teflon sheet and allowed to rest for about 24 hours. The solution splits into two phases due to settling of TiO$_2$ particles, but stirring with a clean paintbrush results in a viscous, white paint. Two coats of this emulsion was applied onto aluminium substrates using the paintbrush. Curing can be done at room or elevated temperatures. It takes about 48 hours at room temperature whereas the process is complete in 60 minutes at 80~$^\circ$C. 

A field emission scanning electron microscope (Zeiss Ultra 55) was used to image and characterise the titania powder particles. Reflectance was measured in the ultraviolet, visible and near-infrared spectra (300--2500~nm) using PerkinElmer Lambda 950 UV-Vis-NIR spectrometer. Absorbance of the coating in the frequency band of 4000--400~cm$^{-1}$ ($\equiv$ 2.5--25~$\mu$m) was characterised using the Hyperion 3000 infrared microscope augmented to the Bruker Vertex 80, FTIR spectrometer. Coating thickness was measured by Bruker DektakXT stylus profiler.

\subsection{Results}

SEM analysis of TiO$_2$ powder particles (\cref{fig:SEM_supp}) reveal that the particles follow a normal size distribution, with the mean particle radius at 125~nm. However, for all further simulations a uniform particle size distribution shall be considered for reasons discussed in \cref{subsec:Params_SizeDistri}.  

\begin{figure}[htb]
    \centering
    
    \subfloat[Reflectance]{\includegraphics[width=0.475\textwidth]{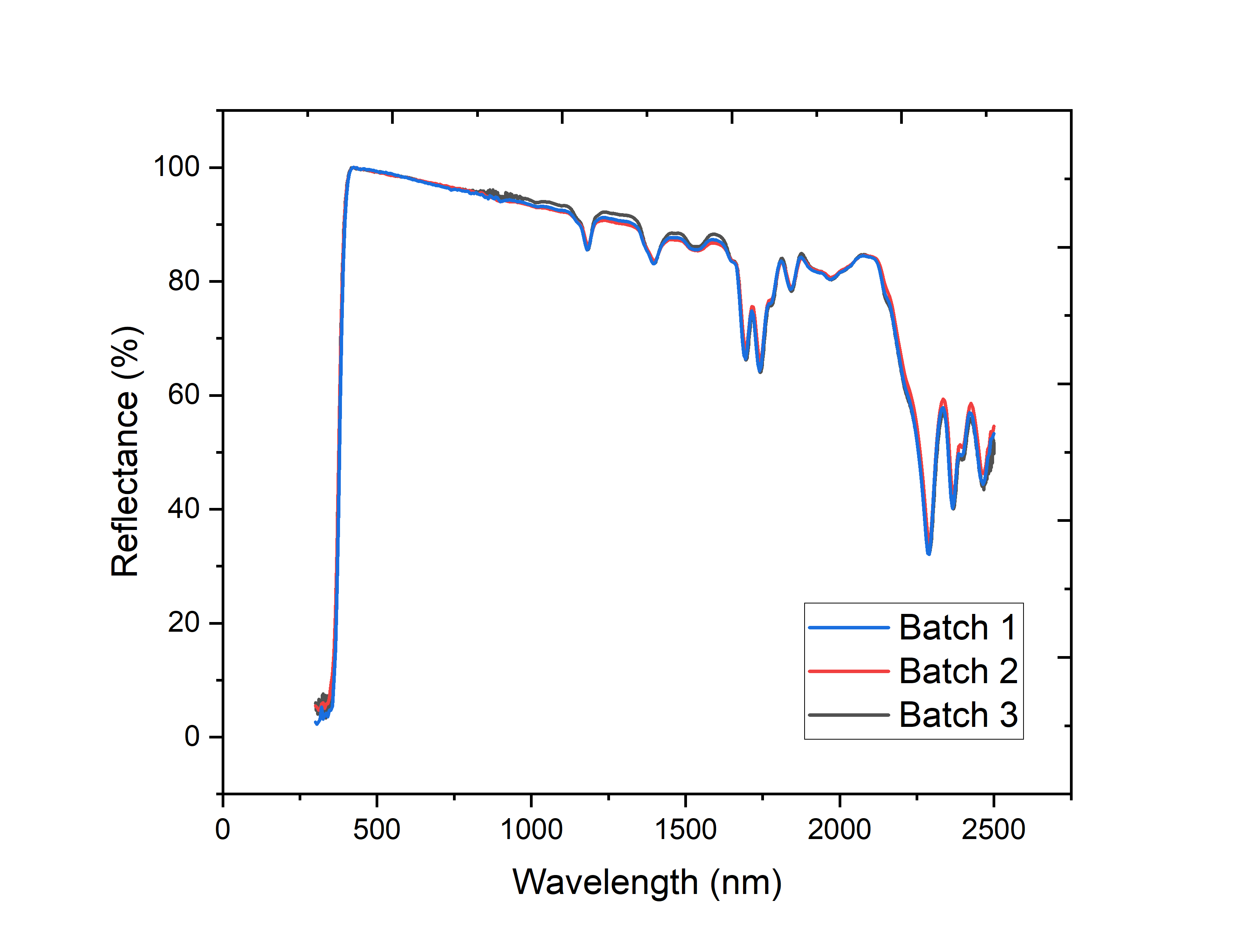}\label{fig:BatchRCompare}} 
    \subfloat[Absorbance]{\includegraphics[width=0.475\textwidth]{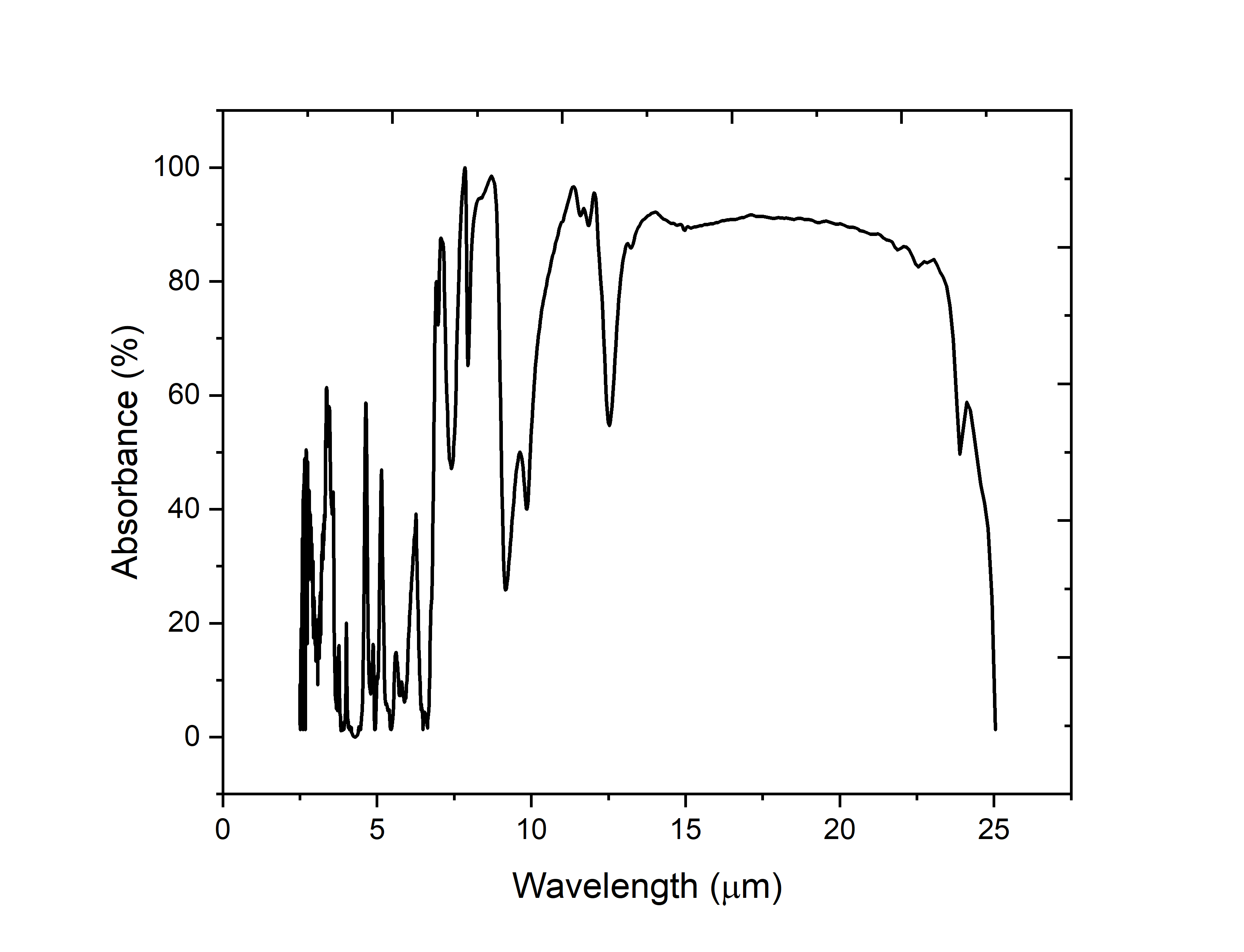}\label{fig:Coat_Abs}}
    
    \caption{Spectral reflectance and absorbance of fabricated coating}
    \label{fig:Refl_Abs}
    
\end{figure}

Multiple batches of the paint were prepared, coated on aluminium substrates and tested for their spectral characteristics. Reflectance measurements (\cref{fig:BatchRCompare}) of the fabricated paint give an average solar reflectance of 91.6~\%. This high solar reflectivity translates to a significant reduction in heat gains under direct insolation. A low intake of energy from solar irradiation consequently lowers the roof surface temperature -- which otherwise records  temperatures higher than that of ambient air. This directly reduces the ceiling temperatures recorded, demonstrating a decrease in the need for cooling. Similar behaviour from multiple batches of the coating, as shown in \cref{fig:BatchRCompare}, indicates its reproducibility.

Kirchoff's law equates a body's emissivity to its absorptivity provided it is in thermal equilibrium. The absorbance of the fabricated coating measured using FTIR at ambient temperature, depicted in \cref{fig:Coat_Abs}, can thus be used to study the coating's emissive characteristics. The coating exhibits an emissivity of 75.2~\% in the ATW\footnote{ATW -- atmospheric transparency window $\equiv$ 8-13~$\mu$m}. The prominent dips observed in this regime, around 9, 11 and 12.5~$\mu$m, are due to an increased extinction coefficient (complex part of the refractive index) of PDMS which increases the reflectivity\footnote{Absorbance = 1 - Reflectance, as Transmittance = 0} of the coating (see \cref{sec:coat_char}).

\begin{figure}[htb]
    \centering
    
    \subfloat[Reflectance]{\includegraphics[width=0.475\textwidth]{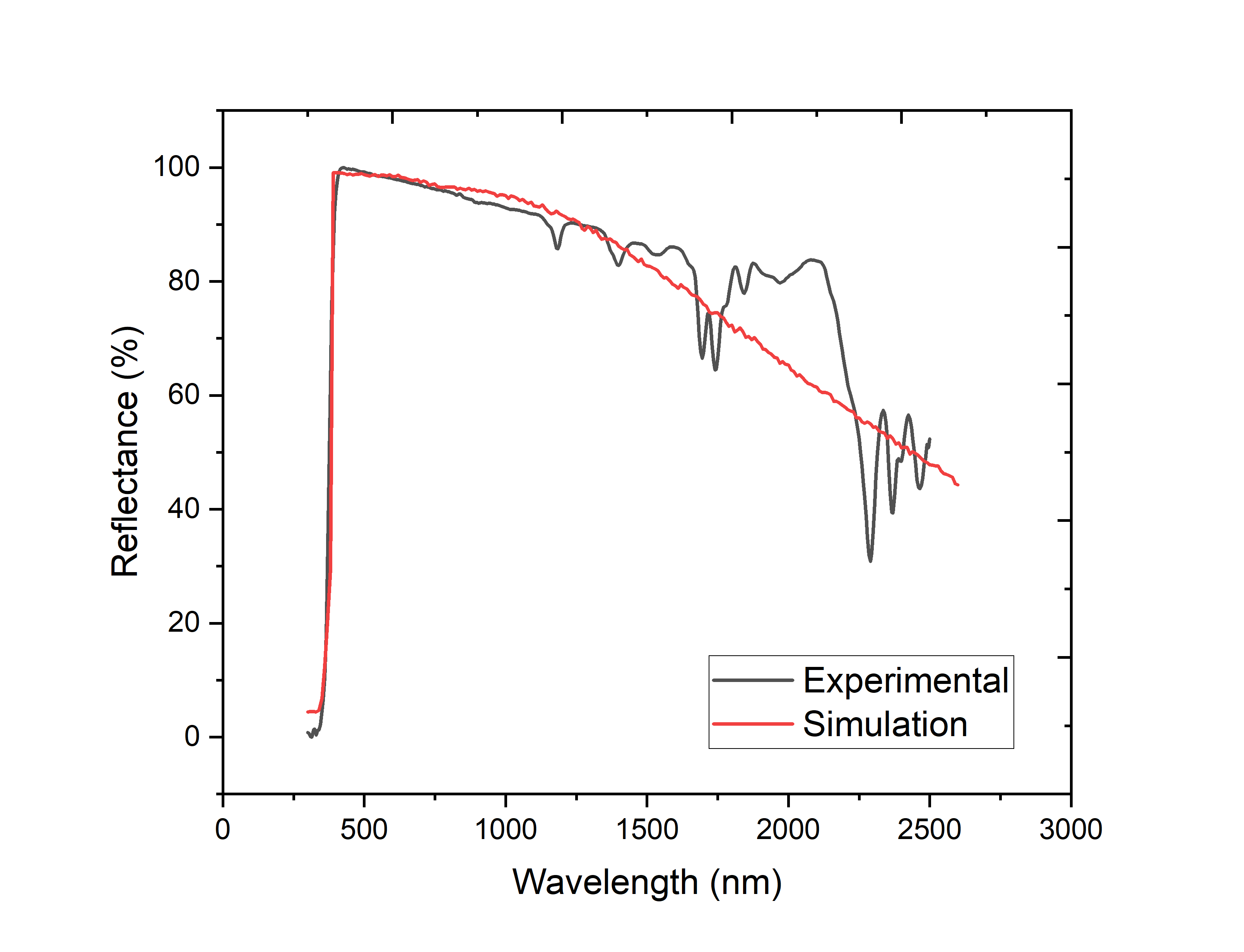}\label{fig:RCompare_Simulation}} 
    \subfloat[Absorbance]{\includegraphics[width=0.475\textwidth]{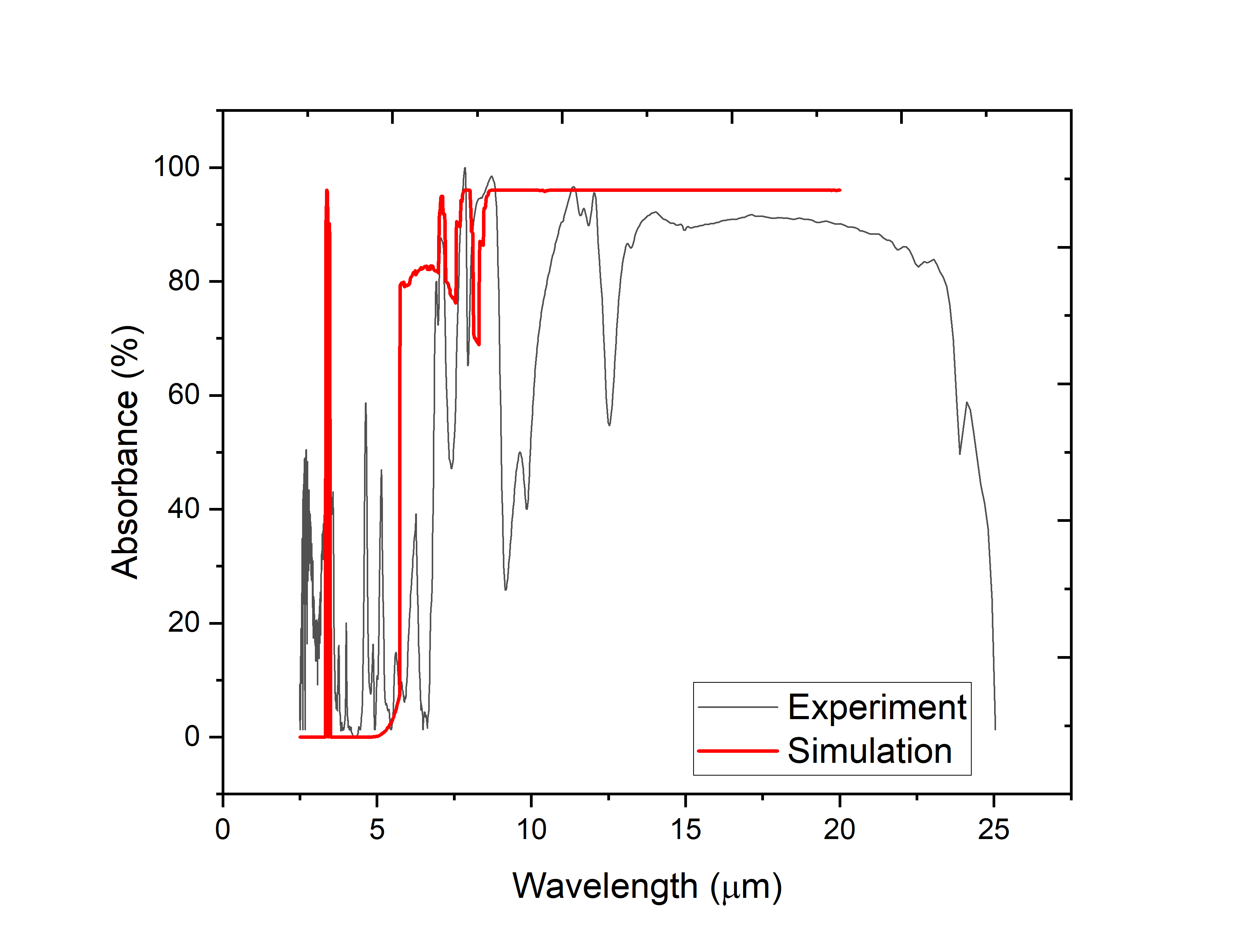}\label{fig:ACompare_Simulation}}
    
    \caption{Validation of theoretical models}
    \label{fig:Expt_Validation}
\end{figure}

Lastly, experimentally obtained results and predictions of our theoretical models are compared to validate the latter. The material system was simulated for the following parameters: a uniform-sized distribution of particles with a radius of 125~nm, occupying a volume fraction of 10~\%, and for a coating thickness of 270~$\mu$m. \Cref{fig:RCompare_Simulation} is proof of good agreement of experimental findings with the Monte Carlo method. \\
A limitation posed by the use of Mie theory is that it does not account for a highly absorbing matrix. The absorption coefficient (obtained from Mie theory) is modified, incorporating matrix absorbance, before being input to the MC algorithm. The simulation result for the same material configuration shows a similar trend as observed in experiments (\cref{fig:ACompare_Simulation}), successfully validating our simulation framework.  

\begin{figure}[htb]
    \centering
    
    \subfloat[Outdoor performance]{\includegraphics[width=0.475\textwidth]{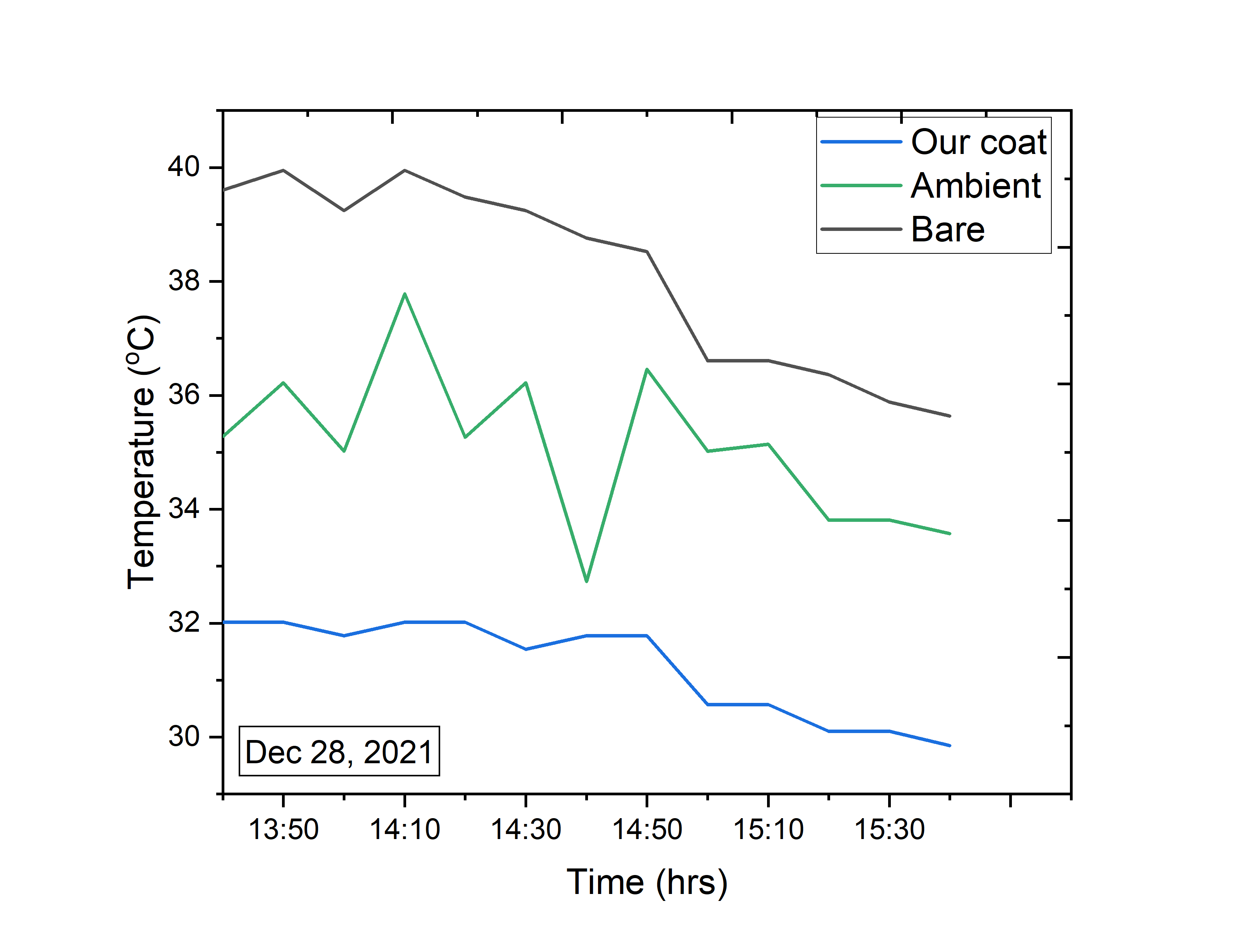}\label{fig:Tm_28Dec2021}} 
    \subfloat[Sub-ambient cooling]{\includegraphics[width=0.475\textwidth]{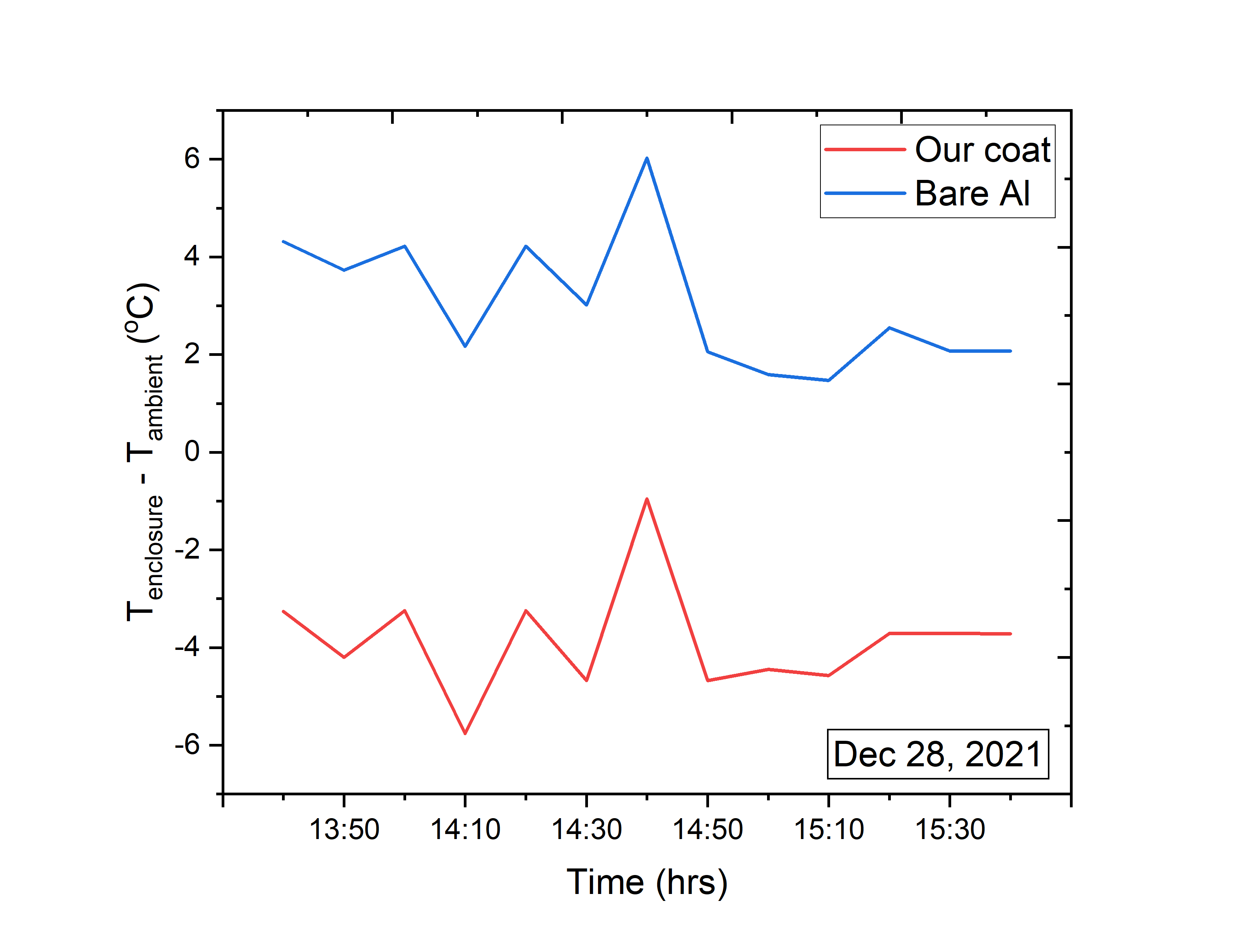}\label{fig:dT_28Dec2021}}
    
    \caption{Testing cooling performance under outdoor conditions}
    \label{fig:Test_Outdoor}
    
\end{figure}

\subsubsection*{Cooling performance}

Cubical aluminium boxes, each of dimensions 6~cm~x~6~cm~x~6~cm, were coated and exposed to the outdoor environment to test for the coating's cooling performance. One box that was left bare acted as the control environment for the experiment. Calibrated thermocouples were plugged into each box -- through a 0.5~cm diameter hole on the ceiling -- to measure the interior space temperature, as shown in \cref{fig:Expt_CoolPerf}, and one to measure the ambient temperature. 

The results from this experiment are illustrated in \cref{fig:Test_Outdoor}. The designed coating is capable to cool the enclosure approximately 3--9~$^\circ$C below ambient, during peak solar irradiation (\cref{fig:dT_28Dec2021,fig:dT_17Nov2021,fig:dT_20Nov2021,fig:dT_9Dec2021}). Studies (referred in \cref{sec:intro}) quote achieving sub-ambient cooling of a surface for a particular solar irradiation. The actual reduction in temperature for an enclosure will be lesser than that for a surface. For instance, \cite{zhang} reports a maximum temperature drop of 10.9~$^\circ$C below ambient of a surface and a 2.1~$^\circ$C difference between inside and outside of their house model. In this study, we report the difference in temperatures measured within a metallic enclosure and the ambient. Thus, comparing cooling performance results of this study with sub-ambient temperature reduction of a surface would not be accurate. Based on the metric of cooling within an enclosure, our coating outperforms materials previously reported to achieve PDRC. 

Furthermore, a non-coated surface would be observed to be at a temperature higher than the ambient which is exemplified by the bare aluminium box in \cref{fig:dT_28Dec2021}. Cooling appliances would thus have to cool from this higher temperature to the lower, chosen temperature in order to attain thermal comfort. The box with our coat sees a temperature reduction of 7--8~$^\circ$C (\cref{fig:dTw-Bare}) from that seen by the non-coated, bare aluminium box. This would decrease the temperature difference that is needed to attain thermal comfort, further reducing the work done by cooling unit compressors. This single-layered, PDMS/TiO$_2$ composite coating thus holds high potential to achieve significant reductions in the cooling demand.

\section{Conclusions}
\label{sec:conclusions}

This study has thus designed, verified and validated the Monte Carlo method to study optical properties of a composite coating. The same was fabricated with experiments demonstrating >91~\% solar reflectivity and >75~\% emissivity in the ATW. This corresponded to the achievement of sub-ambient cooling by 4-9~$^\circ$C under peak solar irradiation. 

More research on this combination of materials to create a composite coating that exhibits passive cooling can further improve its thermo-optical performance. For instance, dampening the absorption of PDMS in the NIR region as well as reducing the reflectivity exhibited in the 8--13~$\mu$m band would enhance the cooling performance.  

A TiO$_2$/PDMS coating can thus help reduce the cooling demand in buildings. It also finds application as coatings over public transport such as buses, trains, etc. This metamaterial coating has the potential to considerably reduce emissions assisting the fight to mitigate the human ecological footprint.

\bibliographystyle{unsrt}


\newpage

\appendix

\section{SEM images: Titania powder}
\label{sec:TiO2_partsize_SEM}

\begin{figure}[htb]
    \centering
    
    \subfloat[]{\includegraphics[width=0.45\textwidth]{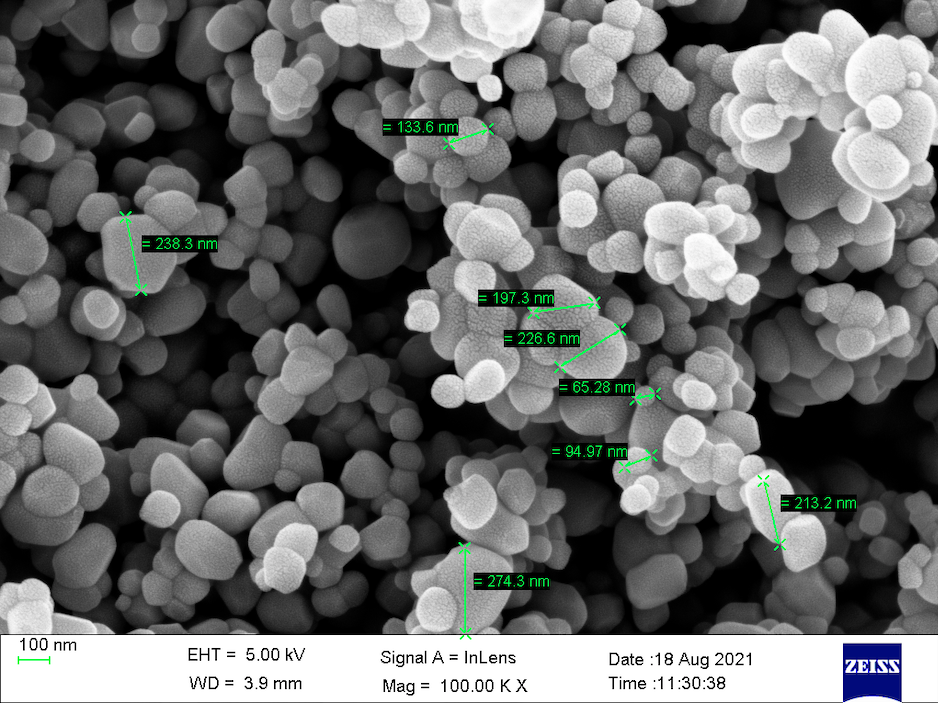}\label{fig:SEM_1}} 
    \subfloat[]{\includegraphics[width = 0.45\textwidth]{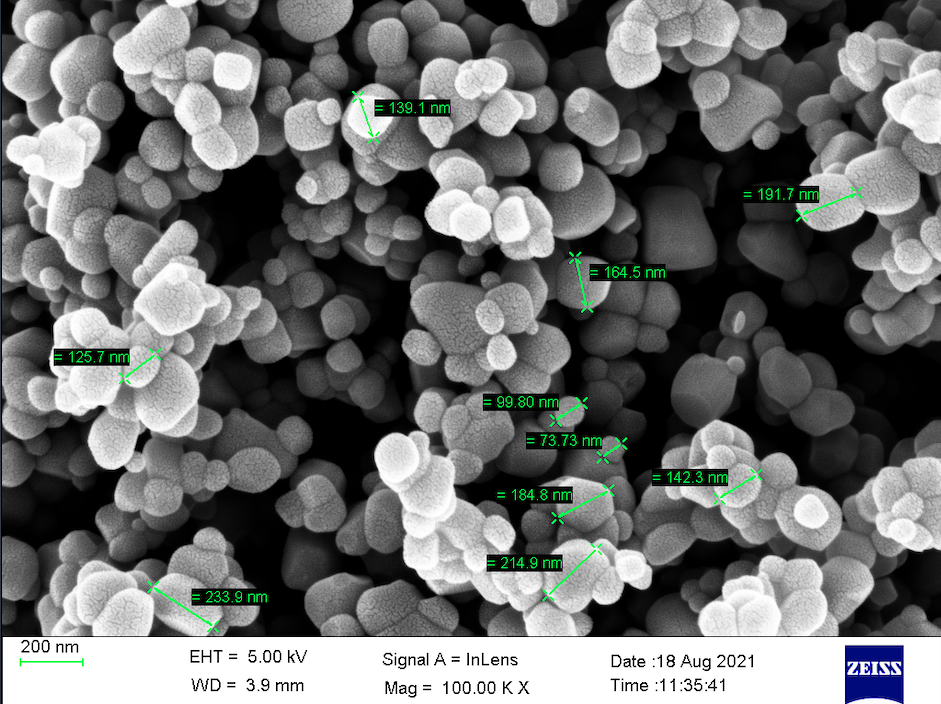} \label{fig:SEM_2}} \\
    
    \subfloat[]{\includegraphics[width=0.45\textwidth]{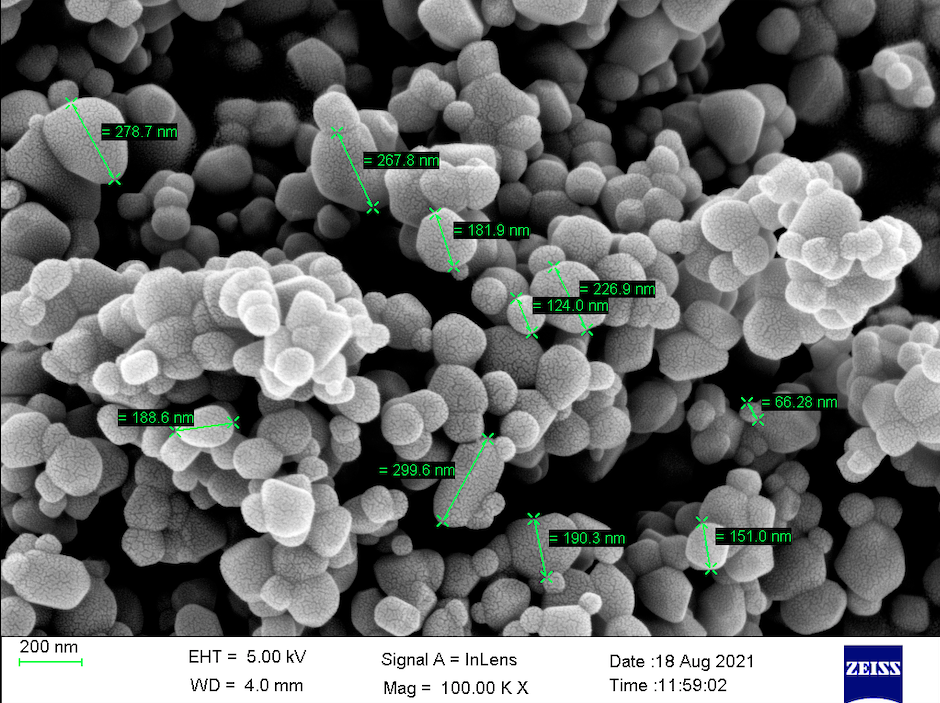}\label{fig:SEM_3}} 
    \subfloat[]{\includegraphics[width = 0.45\textwidth]{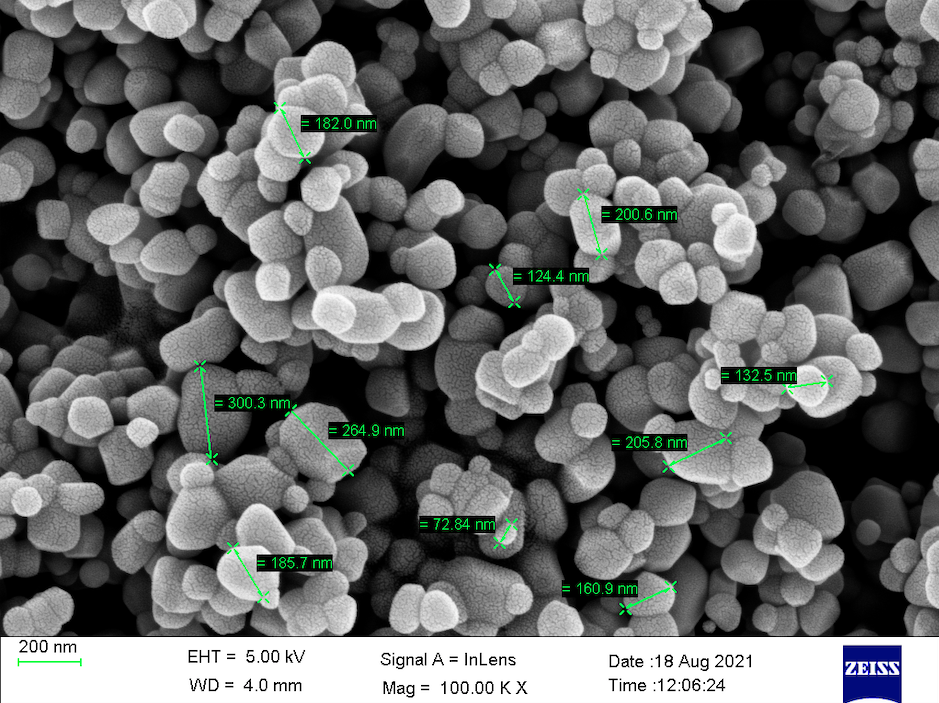} \label{fig:SEM_4}} 
  
    \caption{SEM images of TiO$_2$ powder particles with size measurements.}
    \label{fig:SEM_supp}
    
\end{figure}

Three powder samples were imaged under SEM and the images, shown in \cref{fig:SEM_supp}, were analysed to determine the particle sizes. The particles are observed to follow a distribution of sizes and are not uniform. Assuming a normal distribution of sizes, we estimate that the mean particle radius lies about 125~nm with a standard deviation of 15~nm. 

\subsection{Parameter input to model}
\label{subsec:Params_SizeDistri}

\begin{figure}[h]
    \centering
    \includegraphics[width = 0.6\textwidth]{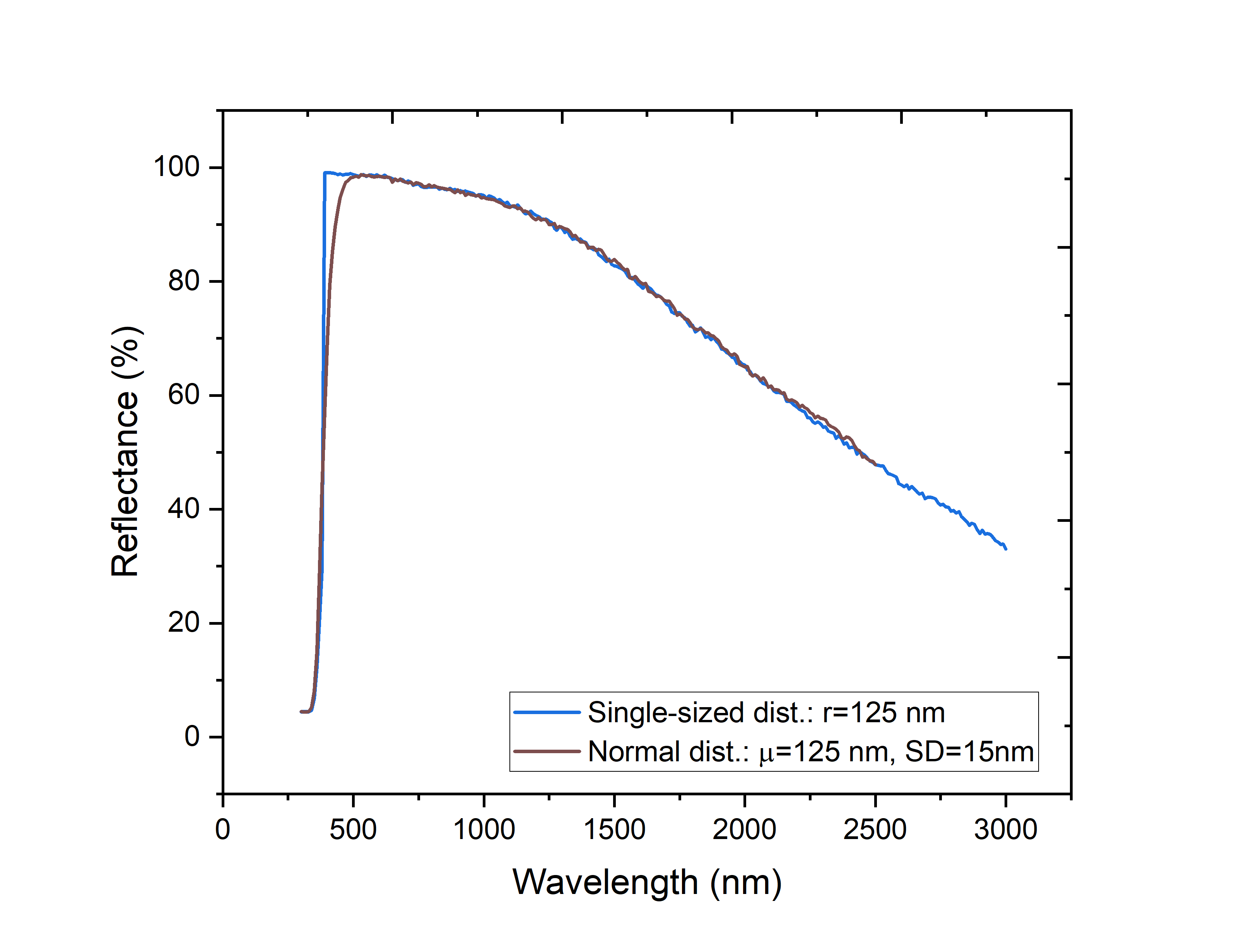}
    \caption{Simulation results for single-sized and normal distribution of particles}
    \label{fig:Uni_vs_Normal_Dist}
\end{figure}

Mie theory only accounts for a single-sized particle distribution. In another work from our research group, codes have been modified to introduce a (normal) distribution of particle sizes to calculate the input parameters to the MC method. Results for both types of distribution were fed into the Monte Carlo simulation to give the solar reflectance as shown in \cref{fig:Uni_vs_Normal_Dist}. The lack of a substantial change of reflectance is thought to be caused by, (a) an increased average size of a scatterer due to agglomeration of individual powder particles, and (b) the particle sizes are insignificant compared to the wavelength of interaction. It is expected that a deviation would be more prominent when individual particle sizes approach dimensions of the wavelength of light (radii about 600~nm). We thus assume a uniform size distribution of the procured TiO$_2$ powder with an average particle size of 250~nm for all further simulations and analyses.

\newpage

\section{Coating characteristics: Solar reflectance and IR emissivity}
\label{sec:coat_char}

\begin{figure}[htb]
    \centering
    
    \subfloat[]{\includegraphics[width = 0.45\textwidth]{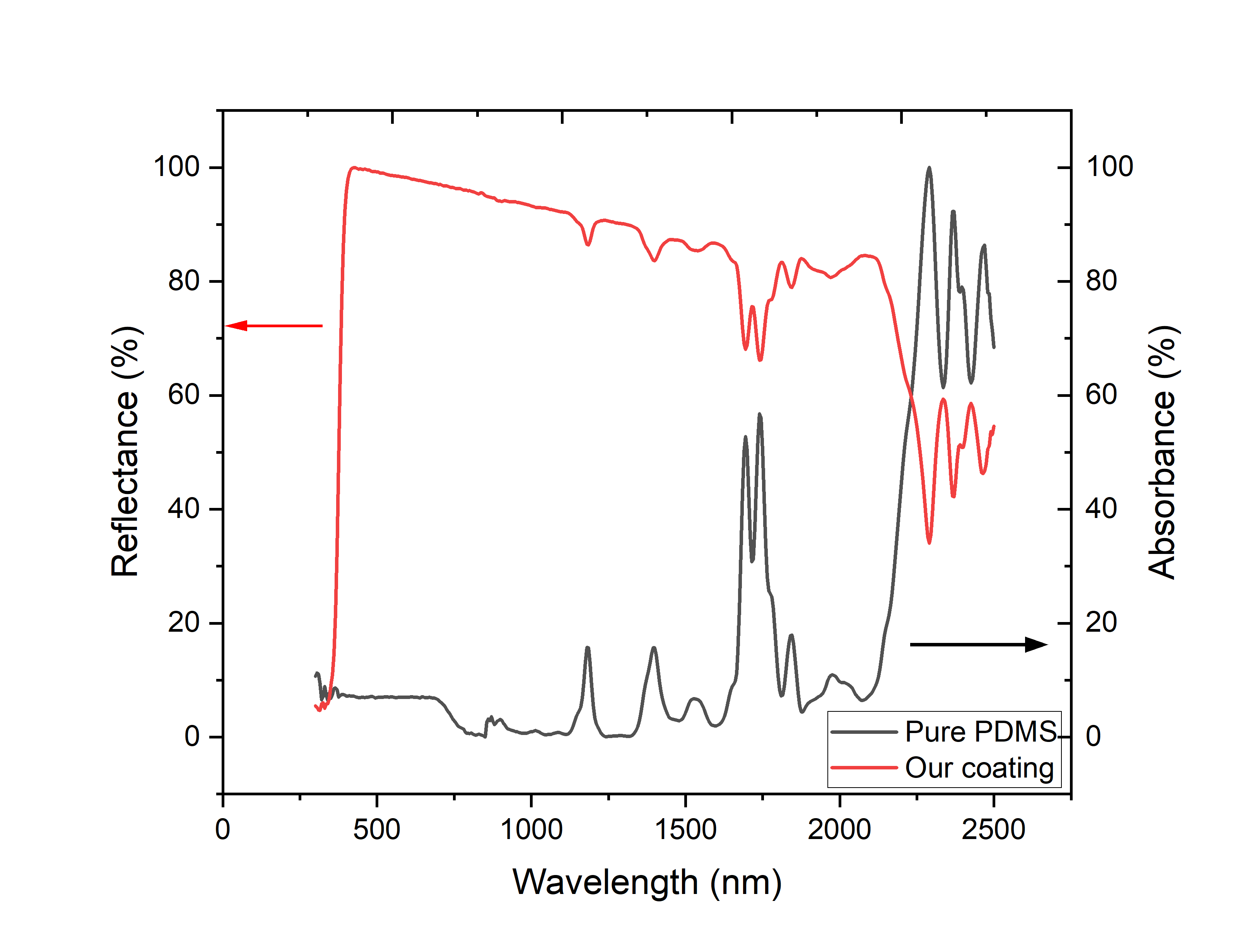} \label{fig:PDMSabs_R}}
    \subfloat[]{\includegraphics[width=0.45\textwidth]{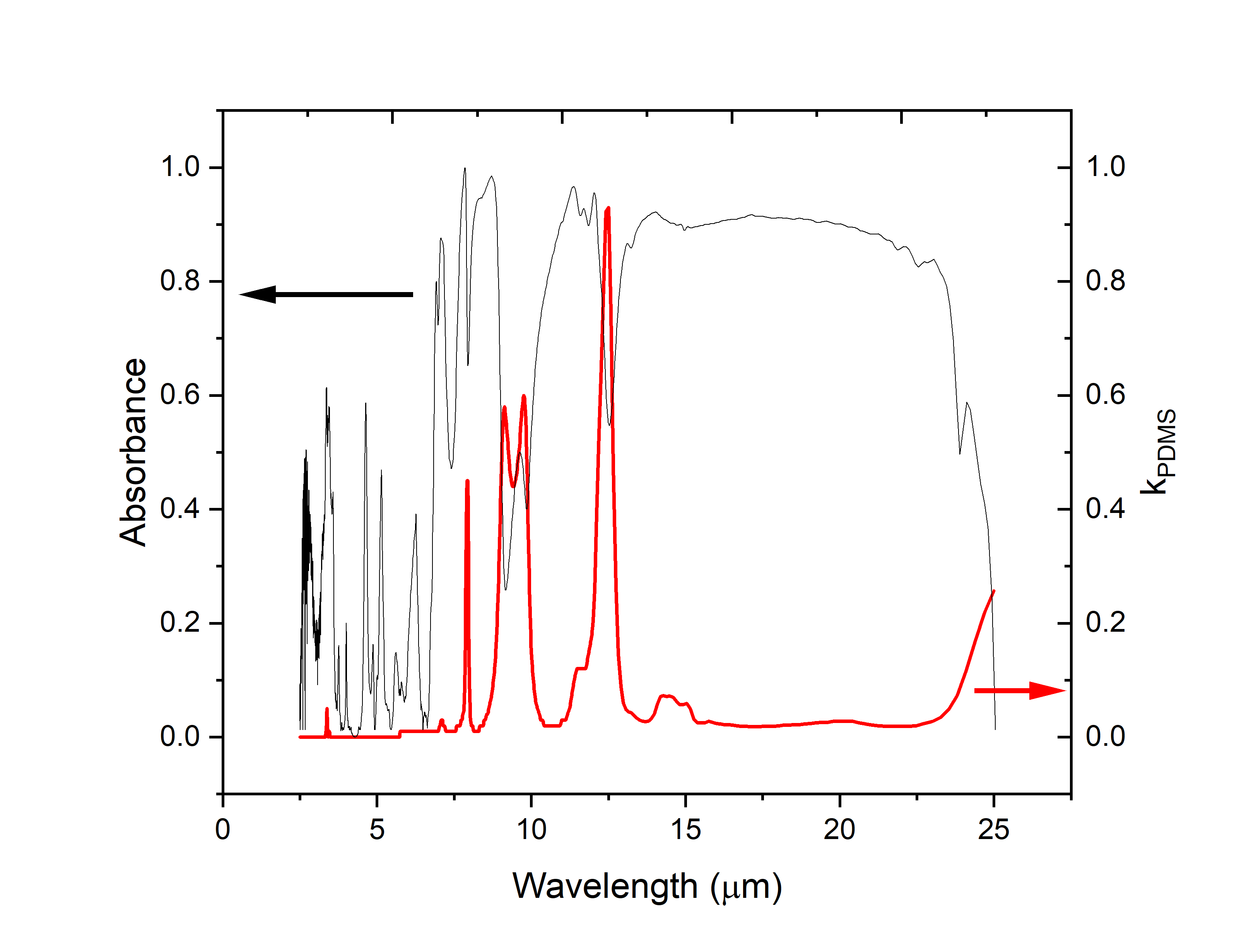}\label{fig:A_k_PDMS}} \\
  
    \caption{(a) Reduction in reflectance due to absorption by the PDMS matrix (b) Reduced absorbance due to higher extinction coefficient of PDMS}
    \label{fig:coat_char}
    
\end{figure}

Dips (or peaks) are observed in the experimentally obtained reflectance spectrum but absent in simulation results. This is a consequence of absorption by the matrix (PDMS), illustrated in \cref{fig:PDMSabs_R}, which is not accounted for by Mie theory and hence is not reflected in the simulations. 

As discussed in \cref{sec:TiO2_partsize_SEM} the type of particle size distribution does not cause considerable change in reflectance properties. So we assume a single-sized distribution with particle radius of 125~nm. TiO$_2$ used during coating synthesis amounts to a volume fraction of 10~\% while the thickness of the coating was confirmed to be 270$~\mu$m by experiment -- employing the stylus profiler.

The absorbance of the fabricated coating, measured using FTIR spectroscopy, shows valley-like features for some wavelengths. On mapping the extinction coefficient (k) of the coating matrix (\cref{fig:A_k_PDMS}), obtained from the refractive index data, it is observed that the valleys in absorbance correspond to the peaks in the k-values. This explains the increased reflectance (hence, reduced absorbance) at those wavelengths.

\newpage
\section{Cooling performance test}
\label{sec:app_CoolPerform}

\begin{figure}[htb]
    \centering
    
    \includegraphics[width=0.5\textwidth]{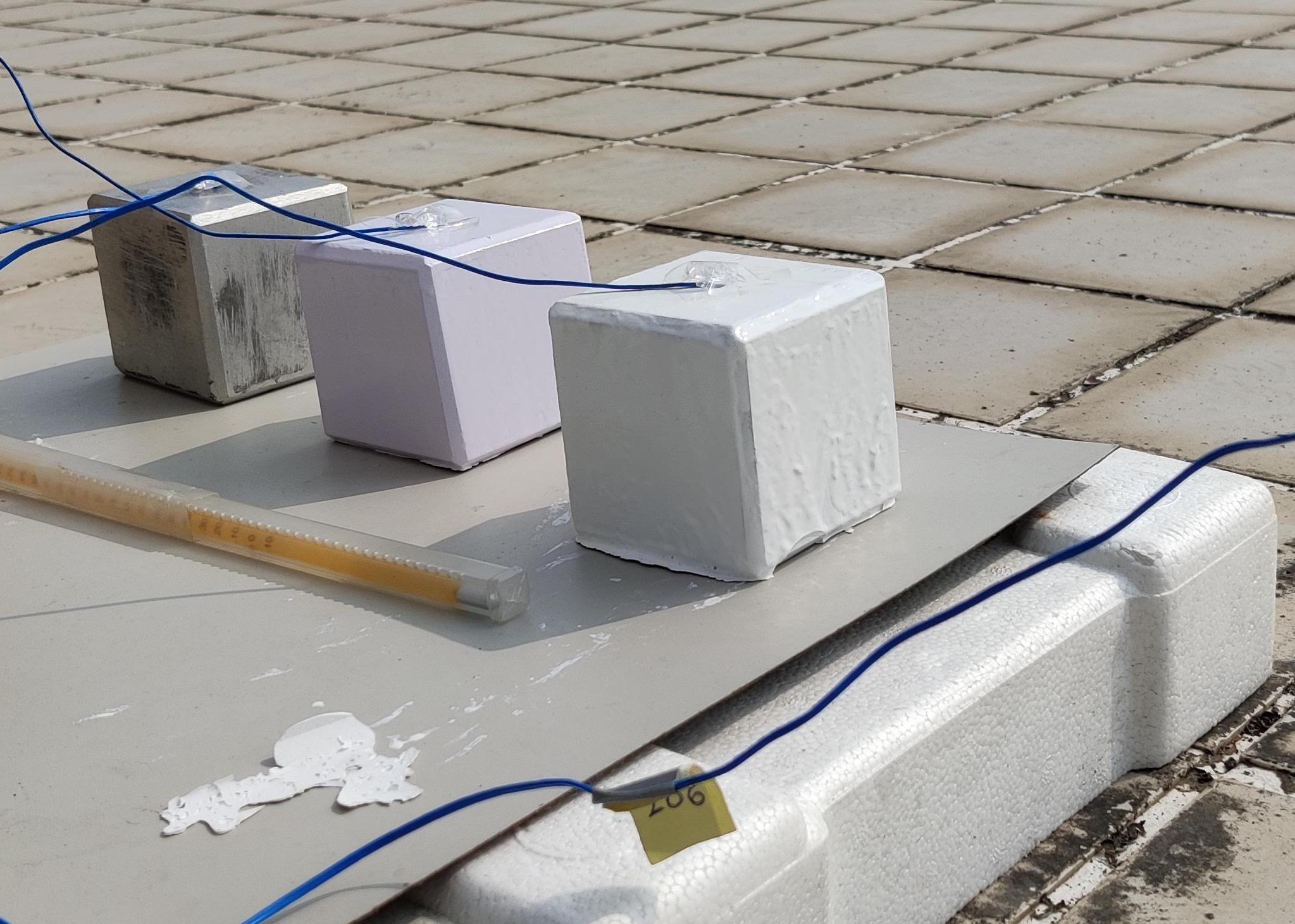}
    \caption{Experimental Setup to measure cooling performance of coatings}
    \label{fig:Expt_CoolPerf}
    
\end{figure}

Three aluminium boxes (cubes of edge 6~cm) are placed on a sheet of mica which, in turn, is placed over rectangular boards of polystyrene (as shown in \cref{fig:Expt_CoolPerf}), limiting the parasitic heat losses. Temperature is measured in regular intervals using calibrated thermocouples. The thermocouples are inserted into the boxes via a hole on the top face and glued to ensure no contact with any face of the box. These tests were conducted on a flat root of a seven-storey building in Mumbai, in November and December 2021.

\begin{figure}[htb]
    \centering
    
    \subfloat[]{\includegraphics[width = 0.45\textwidth]{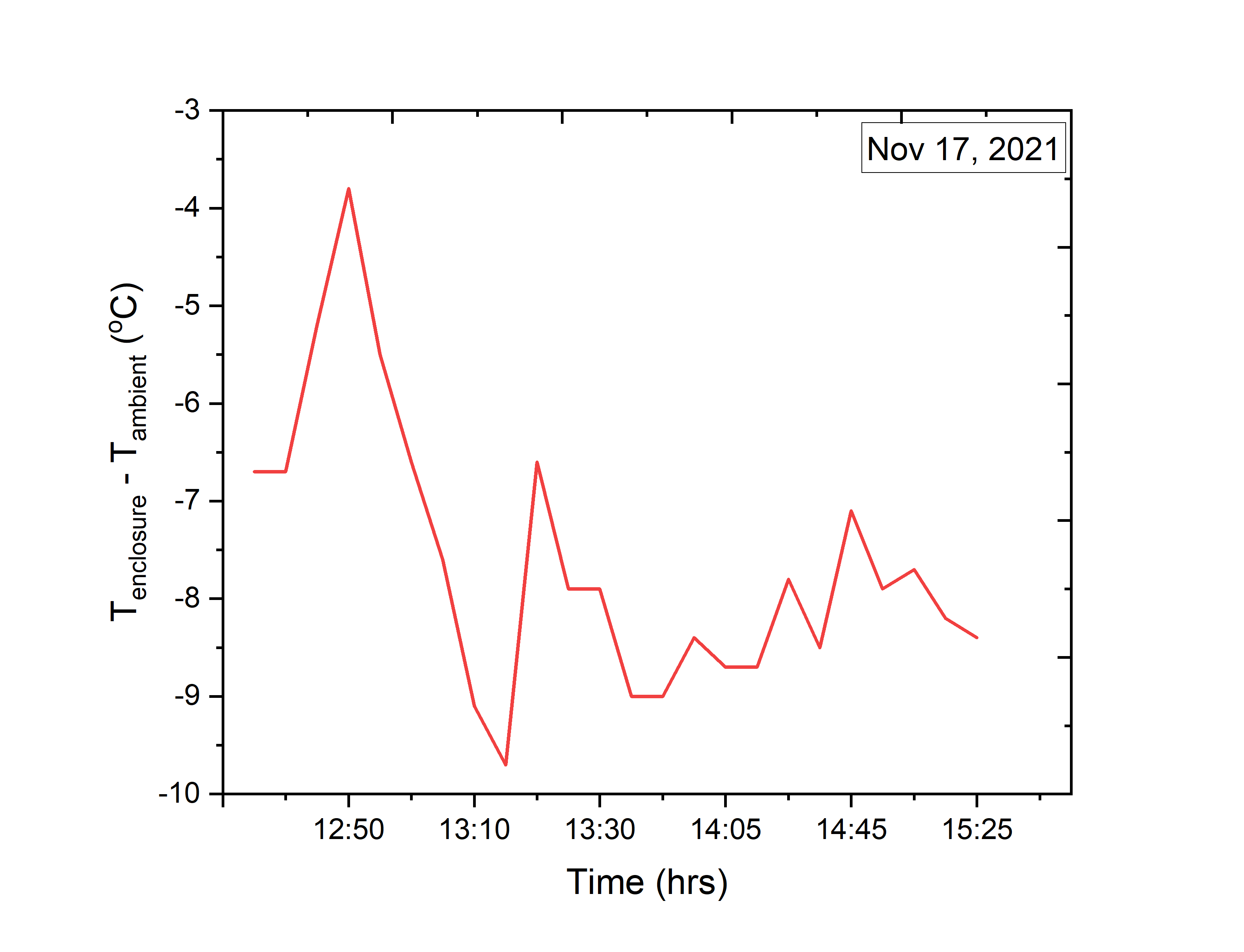} \label{fig:dT_17Nov2021}}
    \subfloat[]{\includegraphics[width=0.45\textwidth]{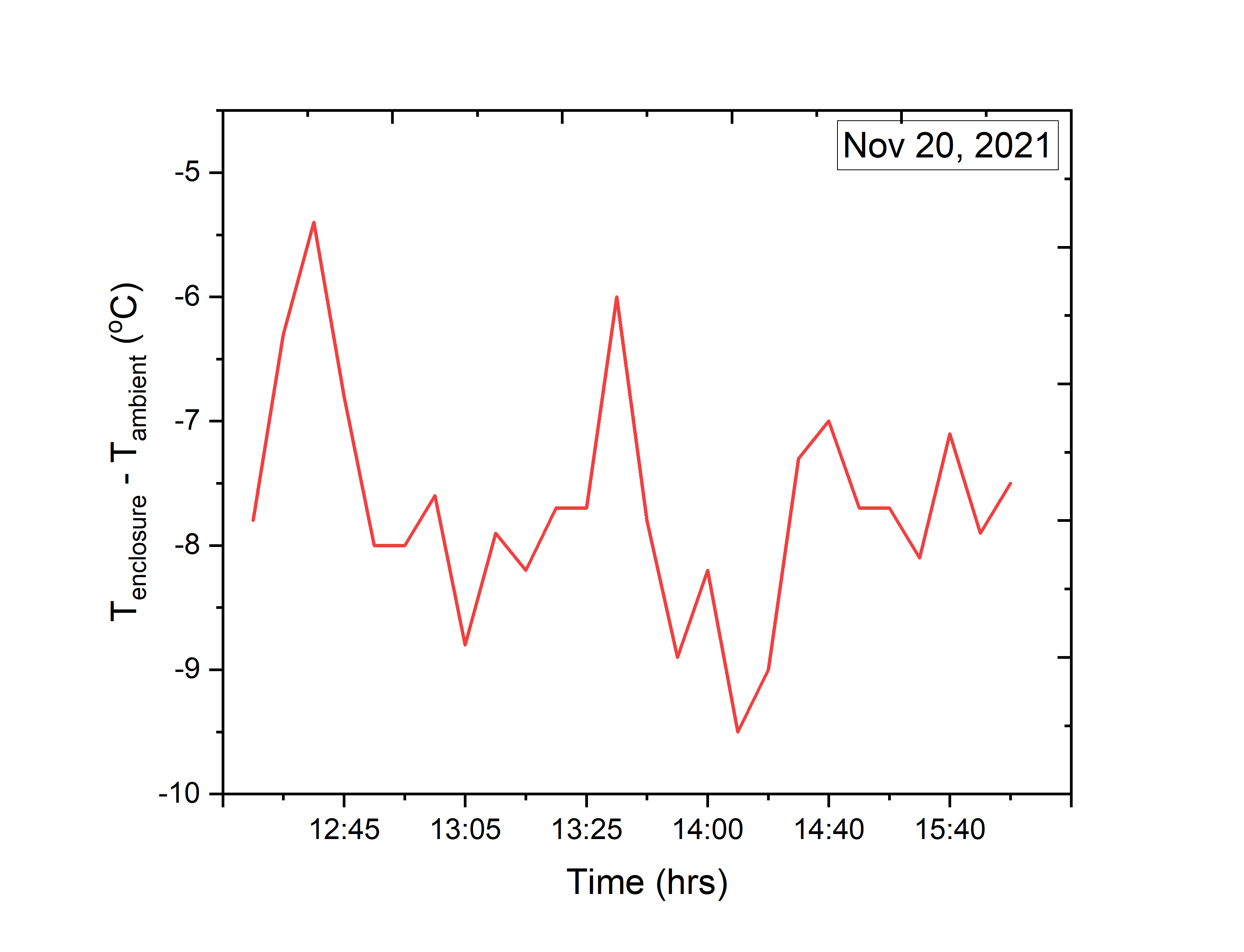}\label{fig:dT_20Nov2021}} \\
    
    \subfloat[]{\includegraphics[width = 0.45\textwidth]{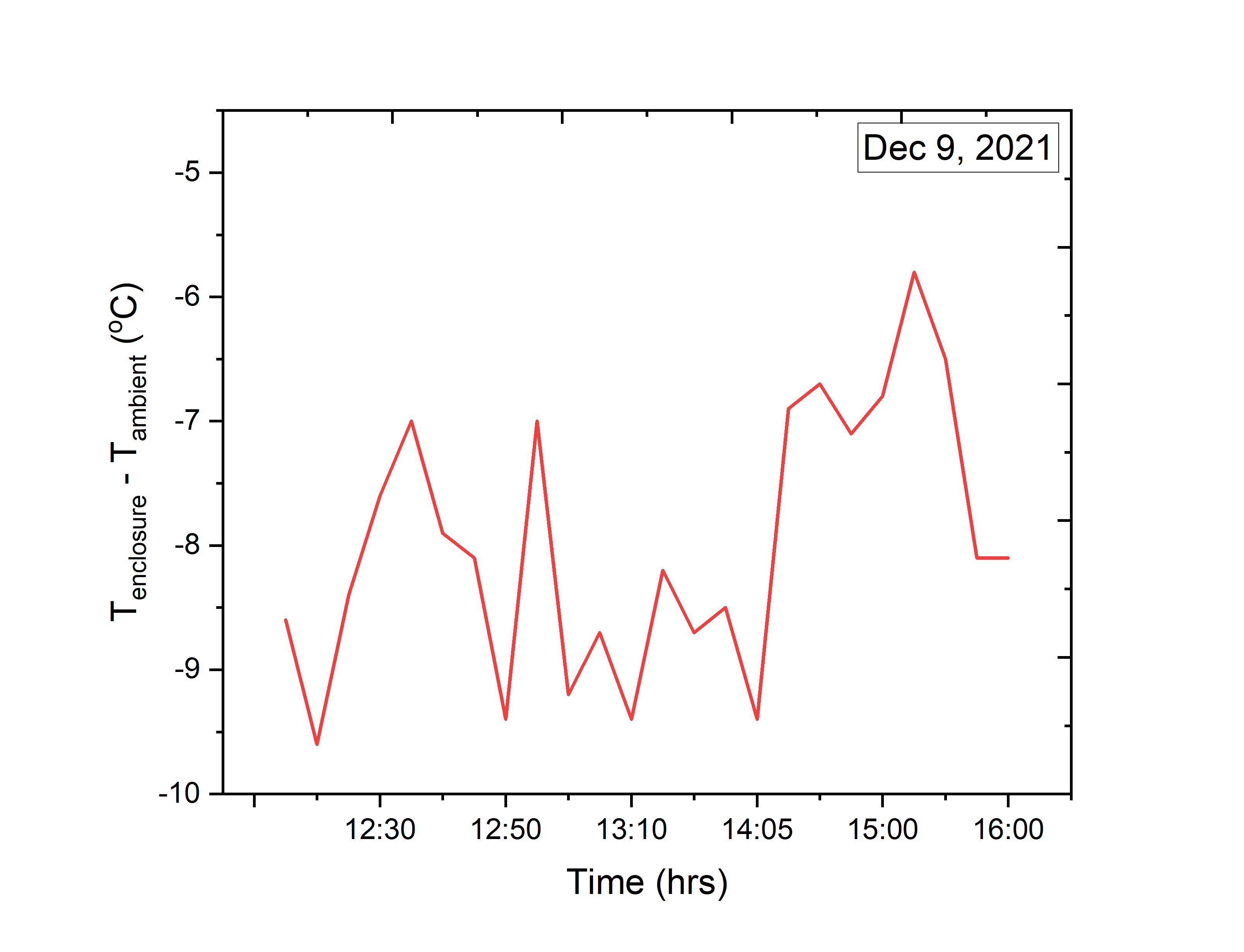} \label{fig:dT_9Dec2021}} 
    \subfloat[]{\includegraphics[width=0.45\textwidth]{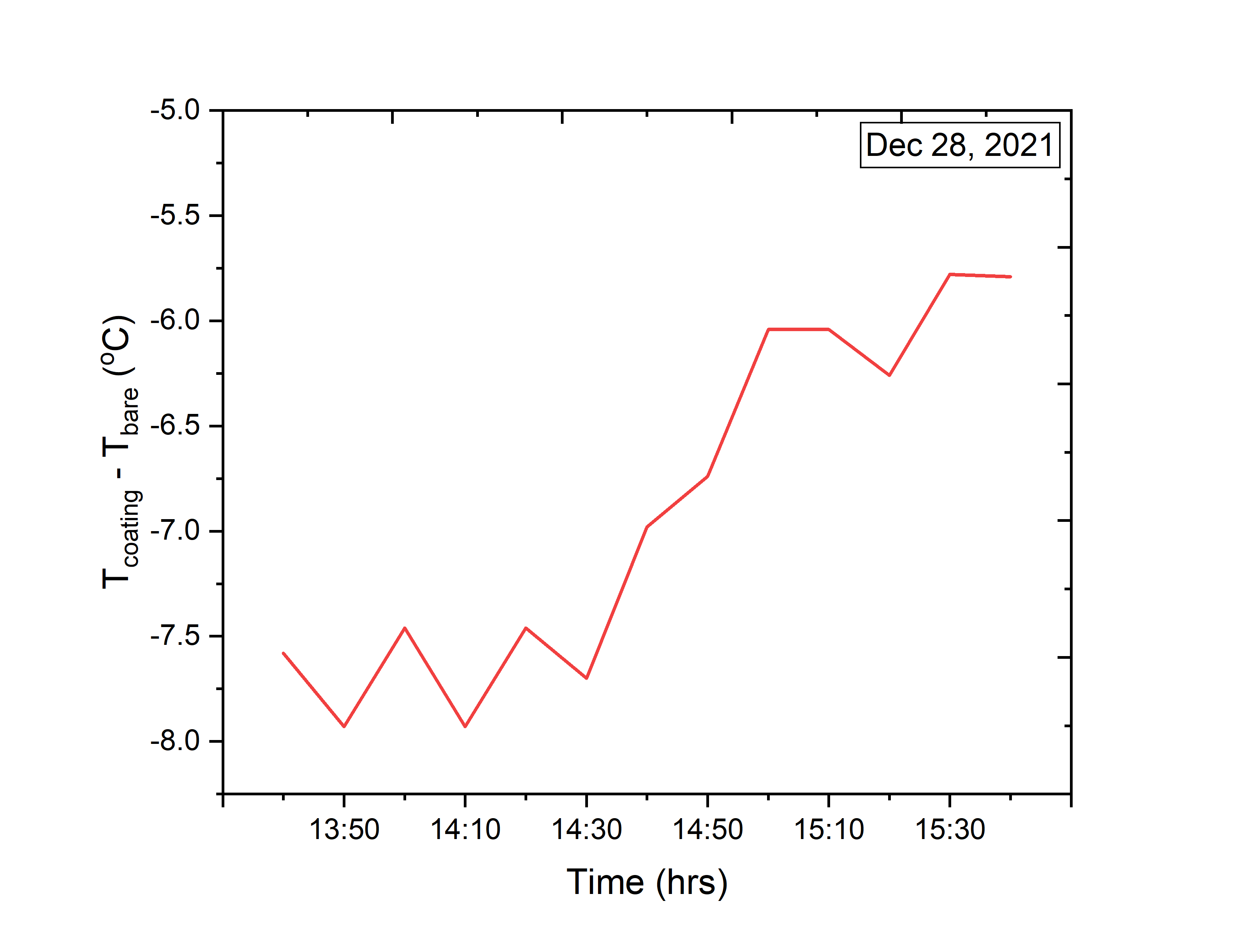}\label{fig:dTw-Bare}}
  
    \caption{(a--c) Quantification of sub-ambient cooling demonstrated on different days of the year (d) $\Delta$T with respect to non-coated Aluminium box}
    \label{fig:dT_expt}
    
\end{figure}

This experiment was repeated on different days, for varying duration of 120--210 minutes. The reduction in temperature from ambient observed on the respective days are illustrated in \cref{fig:dT_17Nov2021,fig:dT_20Nov2021,fig:dT_9Dec2021}. 
The fabricated TiO$_2$/PDMS coating was observed to cool the coated Al enclosure by 5--9$^\circ$C below the ambient temperature, under direct solar irradiation. 

\end{document}